\documentclass[aps,prl,twocolumn,showpacs,groupedaddress]{revtex4}  
\begin{document}
\input epsf
\parskip=0pt
\topsep=0pt
\def\avrg#1{{\langle #1 \rangle}}
\newcommand{\bx}{\mathbf{x}}
\newcommand{\bk}{\mathbf{k}}
\newcommand{\hq}{\hat{q}}
\newcommand{\ie}{{\it i.e.}\,}
\newcommand{\eg}{{\it e.g.},\,}
\newcommand{\fnl}{F_{\text{NL}}}
\newcommand{\chilf}{\chi_{\text{LF}}}
\newcommand{\lsim}{\mbox{\raisebox{-.6ex}{~$\stackrel{<}{\sim}$~}}}
\newcommand{\gsim}{\mbox{\raisebox{-.6ex}{~$\stackrel{>}{\sim}$~}}}
\addtolength{\abovecaptionskip}{-3.5mm}
\addtolength{\belowcaptionskip}{-5.mm}
\addtolength{\abovedisplayskip}{-2mm}
\addtolength{\belowdisplayskip}{-2mm}

\title{\Large\bf  Non-Gaussian Spikes  from   Chaotic Billiards in  Inflation Preheating}
\author{J.~Richard  Bond$^{1}$, Andrei V. Frolov$^{2}$, Zhiqi Huang$^{1}$, and Lev Kofman$^{1}$}
\address{${}^1$ CITA, University of
Toronto, 60 St. George Street, Toronto, ON M5S 1A7, Canada}
\address{${}^2$  Physics Department,  Simon Fraser University, Vancouver, BC V5A 1S6, Canada}

\date{\today}

\begin{abstract}
A new class of non-Gaussian curvature fluctuations $\zeta_{pr} (\bx)  \equiv \delta N(\chi_i) $ arises from  the post-inflation preheating behaviour of a non-inflaton field $\chi_i$. Its billiard-like  chaotic  dynamics  imprints regular log-spaced narrow spikes in the number of preheating e-folds $ N(\chi_i)$. We perform highly accurate lattice simulations of  SUSY-inspired quartic inflaton and coupling potentials in a separate-universe approximation to compute $N(\chi_i)$ as a function of the (nearly  homogeneous) initial condition $\chi_i$. The super-horizon modes of $\chi_i(\bx) $ result in positive spiky excursions in $\zeta_{pr} $ and hence negative gravitational potential fluctuations added to the usual sign-independent inflaton-induced perturbations, observably manifested  in large  cosmic structures and as  (polarized) temperature CMB cold spots. 
 \end{abstract}

\pacs{PACS: 98.80.Cq, CITA-2009-3,  astro-ph/0903.3407}

\maketitle
During early universe inflation, vacuum fluctuations in all light fields $\chi$ are transformed on super-horizon scales into homogeneous and isotropic  Gaussian random fields. These are fully defined by their power spectra $\frac{d\sigma_\chi^2(k)}{d\ln k} \equiv  \avrg{\vert \chi_k \vert^2}\frac{k^3}{2\pi^2}$, whose magnitude, $\sim [H/(2\pi)]^2$, is related to the Hubble parameter $H$ at expansion factor $a$ when $Ha$ first equals the wavenumber $k$ in question.  In standard scenarios, metric curvature fluctuations $\zeta_\phi$ are derived from an inflaton $\phi$ and are nearly Gaussian and often nearly scale invariant,   with power $d\sigma_\zeta^2(k)\approx d\sigma_\phi^2(k) /(2M_P^2 \epsilon)$, in terms of the acceleration parameter  $\epsilon \equiv -\frac{d\ln H}{d\ln a}$ and the reduced Planck mass $M_P \equiv 1/\sqrt{8\pi G_{\text{N}}}$ \cite{fluct}.  Observational constraints on $\epsilon$ are $< 0.03$ at the 95\% confidence limit from the large angle cosmic microwave background radiation (CMB), hence $\zeta$ is considerably amplified over  $\phi$.  Alternative mechanisms utilizing light ``non-inflatons"  $\chi$ must overcome this $\epsilon$ effect. Examples  are  ``curvatons" which temporarily dominate the energy density after inflation \cite{curvaton},
 and  fields which induce spatial  variations of couplings  which modulate the timing of the now-inhomogeneous post-inflation (p)reheating \cite{modulated}, but are not  gravitationally dominant.  
  
The $\zeta$ -source first proposed in \cite{CR} and studied here is nonlinear resonant preheating inducing expansion factor variations from the end of inflation at $\epsilon=1$  to when the equation of state settles to $w=1/3$. A sample framework \cite{GKLS}  for this has a potential  $ V =  \frac{1}{4} \lambda \phi^4  +   \frac{1}{2} g^2\phi^2 \chi^2$, with inflation  driven by the first  and particle creation by the second term. If   $g^2/\lambda $ is tuned to be $\sim {\cal O}(1)$, the non-inflaton $\chi (\bx)$ is light during inflation and  accumulates quantum  fluctuations substantially varying on scales much greater than the Hubble scale at the end of inflation $H_e \sim 10^{-7} M_P$. We note that $g^2/\lambda $=2 corresponds to a SUSY model.  On scales shorter than $ H_e^{-1}$, this $\chi (\bx)$ is nearly homogeneous defining ``separate universes" with specified $\chi_i$ as initial conditions for lattice simulations of the fully coupled fields to determine the  nonlinear evolution $w$, which imprints itself on $\ln a (\chi_i )$.  The resulting curvature perturbations $\zeta_{ph} (\bx )$ are differences in the number of e-folds on uniform Hubble (\ie uniform energy density) time hypersurfaces, $\zeta\equiv \delta \ln a \vert_H=\delta N(\chi_i)$ \cite{deltaN}.  If such variations exist, they would be in addition to
 the standard ones.
\begin{figure*}
\vspace{-0.2cm}
\centerline{\epsfxsize=\textwidth\epsfbox{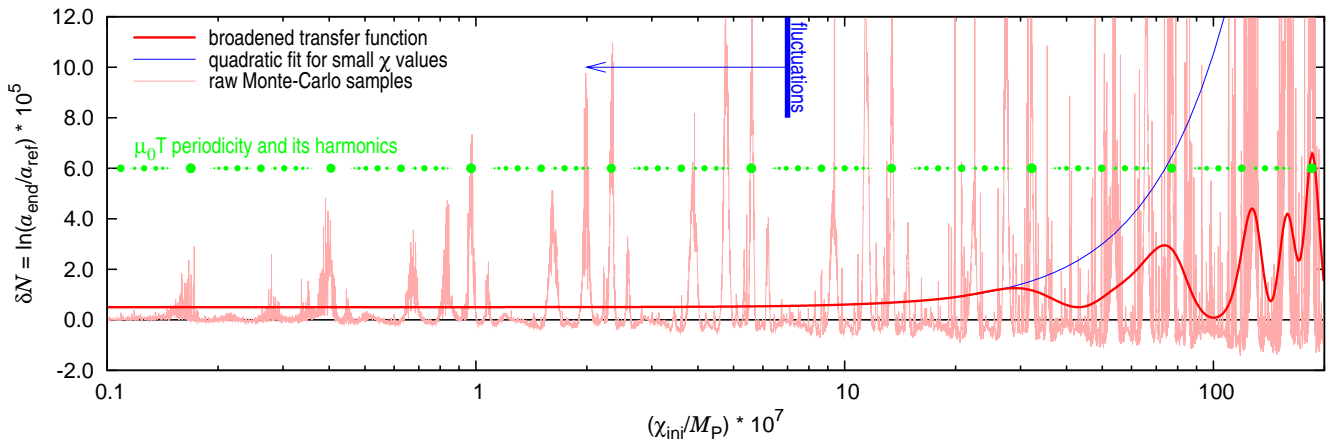}}
\caption{The structure of $\delta N(\chi_i)$ on uniform Hubble hypersurfaces probed with $\sim 10^4$ lattice simulations from the end-of-inflation through the end-of-preheating for varying homogeneous $\chi_i$ initial conditions, for $g^2/\lambda = 2$. The periods $n \mu_0 T$ in 
$\ln \chi_i$ are marked by the large green circles, and the higher harmonics (revealed by the Fourier analysis)
 by smaller green circles. These locate the  spikes in $\delta N(\chi_i)$. The effective field $\avrg{\fnl\vert \chi_b+\chi_{>h} }$ marginalized over high spatial frequencies  with $\sigma_{\text{HF}} $=$7\times 10^{-7}M_P$ (vertical line) yields the  solid curve. A quadratic fit, $f_{\chi}( \chi_b+\chi_{>h})^2$, is also shown. An issue  for {\it our} Hubble patch is whether the ultra-large scale $\chi_{>h}$ is large enough that the large scale structure fluctuations about it, $\pm \sigma_{b<h}$, encompass smoothed peaks in field space, or not. A typical value for $\sigma_{b<h}$ is $\sim 3\times 10^{-7}M_P$. 
}
\label{Fig:numerics}
\end{figure*}
Preheating in the model begins with parametric resonance amplifying the fluctuations  $\chi_k(t) e^{i\vec{k}\cdot\vec{x}}$ describing vacuum excitations  of $\chi$-particles. 
The problem can be reduced   to a flat space-time model by a conformal transformations, 
$\chi_k(t)$ obeys an oscillator equation 
with a periodic  frequency controlled by the
 background  inflaton oscillations $\phi(t)$ 
with (conformal time) period $T \approx 7.416$ \cite{GKLS}. In the resonant bands, $\chi_k(t) \sim  e^{\mu_k t}$ is unstable, an exponent $\mu_k$ is  a function   of $g^2/\lambda$ and $k$. If $g^2/\lambda \sim 2$, 
 the maximum of $\mu_k$ in the first resonant band is located close to $k=0$. Hence a nearly homogeneous $\chi_i$  will be  exponentially unstable in each separate universe, and soon becomes entangled in the complicated mode-mode dynamics of preheating driven by the back-reaction effects of copiously produced $\chi$ and $\delta \phi$ particles. To determine  
 $\delta N(\chi)$  at the part per million level, we need  non-linear lattice simulations with energy conservation accuracy (ECA) well below this. The early attempt  by \cite{CR} used  first-order lattice simulations with  $\mathrm{ECA} \sim 10^{-4}$  to claim an effect at the level  $\delta N \sim 10^{-3}$, with a dominantly-quadratic form $\delta N \approx f_\chi \chi_i^2$ characterized by a constant $f_\chi$ similar to $f_{\text{NL}}$ used in non-gaussianity (nG) studies of the CMB sky \cite{coldspot}.
We  do not confirm the large quadratic {\text{nG}} of \cite{CR}, but do find a nontrivial   $\delta N(\chi_i)$ with a regular sharp-spiked pattern
at the observationally interesting $\sim 10^{-5}$ level,  as shown in Fig.~\ref{Fig:numerics}, with a radically different impact on the sky than the   $f_{\text{NL}}$ story. Although our discovery of the spiked  $\delta N(\chi_i)$ function was a truly numerical one, a posteriori we can explain it by a combination of chaotic zero-mode dynamics  after  $t_e$ and
the abrupt onset of  inhomogeneous nonlinearity at a preheating time $t_{pr}$, allowing us to  conjecture for which  models it works.

To accurately compute the very low levels of $\delta N(\chi_i)$, especially since we are in
 finding  and validation mode for such tiny effects, we needed to go well beyond the ECA practically achievable in second-order preheating codes (typically $\sim 10^{-3}$ for LATTICEEASY \cite{FT} and $\sim 10^{-5}$ for DEFROST \cite{Fr}). We developed a new high-order symplectic PDE solver with adaptive time steps, which can reach machine precision levels (ECA $\sim 10^{-13}$!) to address this problem. In retrospect, we find that ECA of $10^{-7}$ would suffice, which is achievable with shorter timesteps in DEFROST.
The lattice simulations begin at expansion factor  $a_e$  when $\epsilon=1$;
we have shown that variations of the start do not effect results. The calculations are stopped at a uniform Hubble value $H_f$ when the  $w=p/\rho$ is nearly $1/3$, the radiation-dominated value. Although the average $w$ is  $\approx 1/3$, it has small  oscillations during preheating, leading to $a$ fluctuations to average over to determine $N(\chi_i)$ for use in the curvature formula. To deal with this, we used an accurate extrapolation template for the averaged $a(t)$ or, for superb accuracy, a  Kaiser window filter to suppress high frequency oscillations in $a(t)$ by a factor $10^{-8}$. With either, we showed that provided we calculate for 5-6 e-folds after $a_e$, the specific  $H_f$   is not 
important. An
accuracy test was to  show $\delta N$ is effectively zero ($\ll10^{-6}$) and not modulated by $\chi_i$ for  $g^2/\lambda$ out of the resonant band, \eg  at  $g^2/\lambda=1$ and $ 3$, for which  $\mu_0=0$. 
 With the symplectic code,  we ran  a  $256^3$ lattice with a box size $64/H_e$ to check that our conclusions derived using a very large number of  lower resolution simulations using  DEFROST  to build up statistics in $\chi_i$ are accurate.
 The essential effects can indeed  be captured with lower resolution and box size. Fig.~\ref{Fig:numerics} showing a spiky pattern in  $\delta N(\chi_i)$ for $g^2/\lambda=2$ was  produced with  $11563$  $32^3$-simulations  with box size $20/H_e$. The amplitude of spikes increases as $\chi_i$ increases. The spikiest pattern  is at $g^2/\lambda=1.875$, with the spikes broadening away from that, finally  disappearing at  the $g^2/\lambda=1$ and $3$   borders.
 
For  Fig.~\ref{Fig:numerics}, we explored a large range in input $\chi_i/M_P$,  from $10^{-8}$ up to $2\times 10^{-5}$.  $\chi_i (\bx)= \chi_{<h}  + \chi_{>h}   $ has  a sub-horizon contribution $\chi_{<h}$ from eigenmodes with wavenumbers between $k_e \sim H_e a_e$  to  the current horizon scale, $k_{h}\sim H_{h}a_{h}$, and  a super-horizon contribution $\chi_{>h}$ with waves from $k_{h}$ to a $k_{min}$ whose value will depend upon the inflation model. The corresponding variances are 
$\sigma_{<h}^2 = \int_{ k_h}^{k_{e}} d\sigma_{\chi}^2 $ and $\sigma_{>h}^2 = \int_{ k_{min}}^{k_{h}} d\sigma_{\chi}^2 $.   Over $> 100$ e-folds, ${d\sigma_\chi^2}$ is nearly constant for $1< g^2/\lambda < 2$, drops substantially as $a\rightarrow a_e$ for  $g^2/\lambda =0$ ($\sim \phi^4$ as expected), and actually rises for $g^2/\lambda =3$. 
 The least number of e-folds $ \ln a_{h} /a_e$ must exceed  $\sim 50-60$,
 and since $H>H_e$ during inflation, $\sigma_{<h}^2\gsim \ln k_e/k_h$ $[H_e/(2\pi)]^2$ gives a $\chi_{<h}$ enhanced over $H_e$ by  $\sim \sqrt{55}$.  The super-horizon power, $\sigma_{>h}^2 \gsim \ln k_{h}/k_{min}$  $[H_e/(2\pi)]^2$ is also log-enhanced,  and considerably so in our illustrative $\lambda \phi^4$  example.  Thus,  the log factors give larger $\chi_i$, including  a  $\chi_{>h} $  random number which is nearly constant within our Hubble patch, but has a $\sim \pm \sigma_{>h} $  patch-to-patch ``cosmic variance".

\begin{figure}
\centerline{\epsfxsize=0.4\textwidth\epsfbox{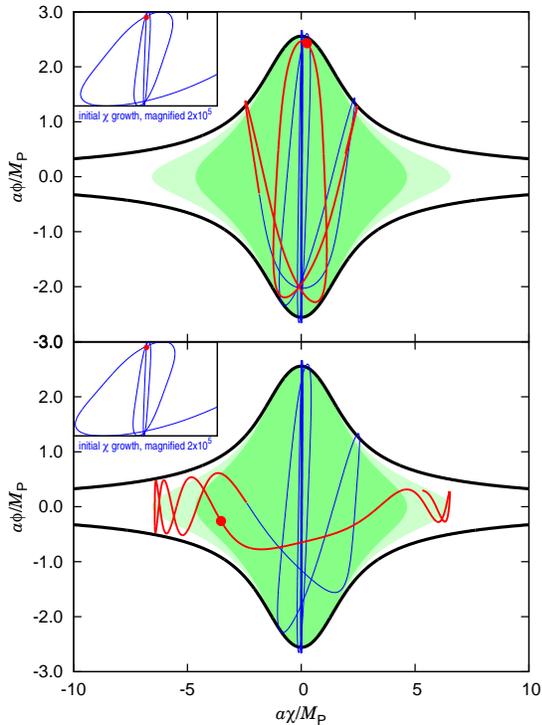}}
\vspace {-0.2cm}
\caption{Billiard trajectories of the $k=0$ modes $\phi(t)$ and $\chi(t)$ within the $\lambda \phi^4/4+g^2\chi^2\phi^2/2$ potential well. Upper panel has a ``no-spike'' initial value
 $\chi_i=3.6 \times 10^{-7} M_P$, and the bottom panel has a ``spike" $\chi_i=3.9 \times 10^{-7} M_P$. The solid curves are  the (fuzzy) potential wall without the inclusion of mass terms induced by field nonlinearities; the pale green and brighter  green border areas include the induced masses at the instances $t=10.8T$ and asymptotically. Thin and thick parts of the trajectories denote before after $t=9.7T$, and up to $11.8T$, with the circle on it  at $t=10.8T$. The inserts in the left upper corners of the panels  show the first several periods of linear oscillations.
 }
\label{Fig:billiards}
\end{figure}

We now show how the features of $\delta N(\chi_i)$ can be understood qualitatively from trajectories in the two-dimensional  space of homogeneous
   modes $(\phi(t), \chi(t))$. The excited  inhomogeneous degrees of freedom do back-react on these $k\approx 0$ modes,  but only  later in the  evolution, \eg at $t\sim 10 T$ for $g^2/\lambda =2$. The  $(\phi(t), \chi(t))$ space is effectively bounded by potential  energy barriers  $ \frac{1}{4} \lambda \phi^4 +   \frac{1}{2} g^2\phi^2 \chi^2 =const$, 
as shown in Fig.~\ref{Fig:billiards}. Initially the trajectories  oscillate  mostly in  the $\phi$ direction, with only  very small initial amplitudes in $\chi$, as illustrated in the insets in  Fig.~\ref{Fig:billiards}.
These oscillations are akin to billiard motions  between the  potential walls.
Precession of the initial oscillations causes the $\chi$ amplitude to grow exponentially in a chaotic manner: $\chi(t)=\chi_i \, e^{\Lambda t} $, where $\Lambda$ is the Lyapunov exponent.
This  gives us  new insight on the parametric resonant  $k=0$ solution in terms of  the Lyapunov instability.  For $g^2/\lambda =2$, we find $\Lambda=\mu_0=0.235$. This conjecture also works for  the cases  in which $k=0$ is not in the resonant band; \eg for $g^2/\lambda=3$ we have $\mu_0=0$ and do indeed find periodic trajectories so $\Lambda=0$.

 As $\chi$ grows  beyond the linear regime, the bouncing billiard  experiences the negative curvature of the  potential walls and a bifurcation of the trajectories occurs, with a few entering the arms in between walls, and  most do not. If there were only homogeneous modes the impact  of this would be temporary because eventually all trajectories would be chaotically mixed. However, the  excitation of the inhomogeneous modes result in exponential growth of $\langle \chi^2 \rangle$ and $\langle \delta \phi^2 \rangle$, as  $e^{2 \mu_* t}$ and  $e^{4 \mu_* t}$ respectively, where $\mu_*$ is an effective resonant exponent \cite{GKLS}. These induce enhanced effective masses in the fields, abruptly changing  the potential, with the arms in between walls closing exponentially quickly,  as shown in  Fig.~\ref{Fig:billiards}.
The trajectories which happen to get into the arms before arm-closure evolve very different from those which never get into the arms. The  billiard picture breaks down when  the gradient terms in $ \chi$ and $\delta \phi$ occur, at arm-closure time, but the bifurcation determined by the
initial conditions at the linear stage has already happened.  The two pre-closure classes of  in-arm and not-in-arm trajectories result in different equations of state, and hence a  $\chi_i$-dependent $\delta N (\chi_i)$. In-arm trajectories experience kinetic energy kicks from the closing arms, which translates into a transient decrease in  $w$, inducing a  jump in $\delta N$.
The no-spike trajectories of the upper panel of Fig.~\ref{Fig:billiards} 
are much more numerous  than the rarer sort in the lower panel corresponds that give spikes. The billiard picture predicts the spike pattern as one  periodic in $\ln \chi$ which works extremely well: the same ``spiky'' trajectory labeled by $\chi_i$ is  repeated for initial values $\chi_i \, e^{n \mu_0 T}$ for integer  $n$.
 The origin of the higher harmonics is more complex. Our spike pattern formula works very well for other values of $g^2/\lambda$, with the requisite $\mu_0(g^2/\lambda)$.
  
 Let us denote $\zeta_{pr}$ by $\fnl( \chi_i (\bx ))$.
These are added to the conventional inflaton curvature fluctuations $\zeta_\phi$: $\zeta_{tot} (\bx) \approx \zeta_\phi (\bx ) +
 \fnl(\chi (\bx )) - \avrg{\fnl}$.  (The ensemble average $\avrg{\fnl}$ over all values of $\chi$, is subtracted so $\zeta_{tot}$ fluctuates about zero.) Structure in $\chi_i$ and hence in $\fnl$, exists on a vast range of resolution scales, from $k_{min}$ through $k_h$ to $k_e$.
 Observed large scale structure (LSS), as probed by the CMB, redshift surveys, weak lensing, and
 rare event abundances such as of clusters probe $\sim$10 e-folds
 in $\ln k$ below $k_{h}^{-1}$, and about the same number of e-folds just below this can be probed with  the more uncertain astrophysical observables involving earlier stages in the nonlinear hierarchy, galaxies,  dwarves, and the  ``first stars" within them.  The superhorizon scales  beyond $k_{h}^{-1}$ have no impact  on $\zeta_\phi$-structure, but do have a large impact on $\fnl$ through the specific value $\chi_{>h}$ built from waves $k_{min}<k<k_{h}$.  To explore the astrophysical consequences of $\fnl$, we  marginalize over high frequency components $\chi_{\text{HF}}$ of $\chi\equiv \chi_{\text{HF}}+\chilf$ to form the conditional  non-Gaussian
``effective  field", $\avrg{\fnl\vert \chilf} $, with $\sim$40-50 e-folds of  ``short distance" substructure in $\fnl$ filtered out.  The mean-field, $\avrg{\fnl\vert \chilf} =  \int \fnl(
 \chi)P(\chi \vert \chilf)d\chi$, is a Gaussian-smoothing of $\fnl$ in field-space, via $P(\chi \vert \chilf ) = \frac{\exp[-(\chi-\chilf)^2/(2\sigma_{\text{HF}}^2)]}{\sqrt{2\pi}\sigma_{\text{HF}}}$,
with variance  $\sigma_{\text{HF}}^2 =\avrg{ \chi_{\text{HF}}^2}$. Using it  will give reliable LSS inferences  if the Fourier  transform of the deviation  $\Delta \fnl\equiv \fnl-\avrg{\fnl\vert \chilf} $ is small for $k^{-1} \gg k_{\text{LF}}^{-1}$, the $\chilf$  filter scale. Since $\chilf=\chi_{b}+\chi_{>h}$ contains a spatially varying part $\chi_{b}$ built from waves with $k_{h}<k<k_{\text{LF}} $ and a constant  superhorizon part $\chi_{>h}$.  Which aspects of the spiky patterns of Fig.~\ref{Fig:numerics} that would be realized in our Hubble patch is quite dependent on the luck of our $\chi_{>h}$-draw
 from a Gaussian distribution with variance $\sigma_{>h}^2=\avrg{\chi_{>h}^2}$, in particular whether $\chi_{>h}$ is near a smoothed-peak, or small $\lsim \sigma_{\text{HF}}$. 
   
 The analytic model, $\fnl=\sum_p F_p \exp[-(\chi-\chi_p)^2/(2\gamma_p^2)]$,  approximates each spike of Fig.~\ref{Fig:numerics} with a Gaussian ``line profile"
 of width $\gamma_p$ centered on $\chi_p$, with peak amplitude
 $F_P$ and integrated line strength $u_p= F_p \sqrt{2\pi}\gamma_p$. The $\chi \rightarrow -\chi$ symmetry means that for each peak at $\chi_p$ there is a mirror peak at $-\chi_p$ of the same strength. The conditional
$n$-point moments of $\fnl$, $\avrg{\prod_{i=1}^{n} \fnl(\bx_i ) \vert \chilf}$ are then easily computable Gaussians, with quadratic terms in $\chi_i$-$\chi_p$ linked through the HF 2-point function of $\chi_{\text{HF}}$.  The mean field has $n=1$:
\begin{equation}\label{fnlchib}
\avrg{\fnl\vert \chilf} = \sum_p  U_p e^{{-\frac{\chi_b^2(\bx)}{2{\sigma_{p\text{HF}}^2}}}}\text{cosh}\Big\{ \frac{\chi_{p>h}}{\sigma_{p\text{HF}}^2} \chi_{b}(\bx) \Big\}.  
\end{equation}
Here  $\sigma_{p\text{HF}}^2 $=$ \gamma_p^2 + \sigma_{\text{HF}}^2$ is $\approx  \sigma_{\text{HF}}^2$ for the typical $\gamma_p$ we find and   $U_p$=$\frac{2u_p \gamma_p}{\sigma_{p\text{HF}}}\exp\{-\frac{\chi_{p>h}^2}{2{\sigma_{p\text{HF}}^2}}\}$, with $\chi_{p>h}$=$ \chi_p$-$\chi_{>h}$. The  $\chi_{\text{HF}}$ correlation function dependence of the fluctuation variance $\avrg{\Delta \fnl (\bx_1 )\Delta \fnl (\bx_2 )\vert \chilf}$ precludes an analytic harmonic analysis, but we have investigated this numerically: \eg for a 1-spike version, this $\fnl$ variance is ~1\% of $\avrg{ \fnl\vert \chilf}^2 $ if
it is smoothed on $\sim k_{\text{LF}}^{-1}$, and  $\sim0.1\%$ with $\sim 10 k_{\text{LF}}^{-1}$ smoothing. (We have checked the effective field works even better for  $\fnl$ quadratic and exponential in $\chi$.) The \text{HF}   structure is very relevant to first-object formation, but what actually happens will be model-dependent.  However, \text{HF} will not impinge upon LSS observables, since they are convolved 
with experimentally-determined or theoretically-imposed windows $\gg k_e^{-1}$ in scale. 
\addtolength{\abovecaptionskip}{1mm}
\begin{figure}
\vspace{-0.2cm}
\centerline{\epsfxsize=0.4\textwidth\epsfbox{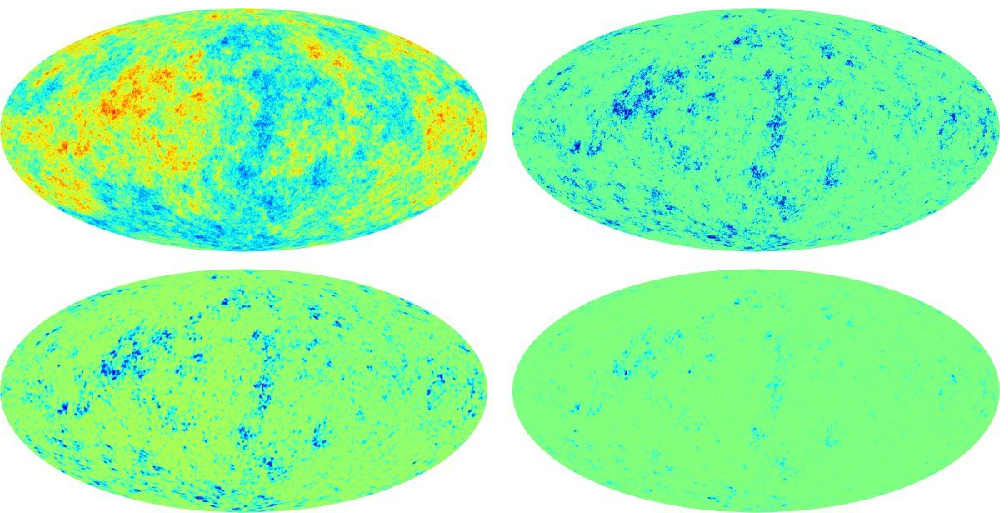}}
\caption{Realizations of the {\text{nG}} map $\avrg{\fnl \vert \chi_b}$ on the CMB sky. Top left  shows a  scale-invariant Gaussian random field realization $\chi_b (d_{\gamma dec} \hq )$ in direction $\hq$ on a sphere at the comoving distance to photon decoupling, $d_{\gamma dec}$  Top right shows the action of   $\avrg{\fnl \vert \chi_b}$ on it, using our Gaussian-line-profile approximation with 2 peaks at $\chi_p=\pm \nu_p \sigma_b$, for $\nu_p =3.5$. Middle left shows the map convolved with a CMB transfer function, and smoothed on a $1^\circ$ scale,   right with  $\nu_p =4.5$; both show ``cold spot"  intermittency. }
\label{Fig:sphere}
\end{figure}
  
A fundamental character of the resonant mechanism is the delay of in-arm preheating completion, translating into positive large excursions in 
$\delta \ln a\vert_H$. The associated perturbed Newtonian gravitational
potential $\Phi_{\text{N}}$ is negative.  The CMB sky temperature $T(\hq)$, an angular function of the CMB photon direction $\hq$ towards us,  is a projected image of various sources whose 3D Fourier transforms involve various form factors ${\cal F}(\vert \bk \vert)$ times  $\Phi_{\text{N}} (\bk , t_0)/3$. The dominant ${\cal F}$ terms are from two CMB decoupling effects, and one late-time effect: the combined ``naive Sachs-Wolfe (NSW) effect plus photon compression-rarefaction"; the Doppler effect from flowing electrons;  and the integrated Sachs-Wolfe (ISW) effect with ${\cal F}$ a $k$-dependent time-integral of $ 6  \dot{\Phi}_{\text{N}}/{\Phi}_{\text{N}}$ \cite{fluct}. The upper panels in Fig.~\ref{Fig:sphere} correspond to ${\cal F}=1$  and the lower panels are convolved with a CMB transfer function and smoothing on $1^\circ$, appropriate if the CMB sky is a direct map of the photon decoupling surface, ignoring its fuzziness (valid for  the $1^\circ$ smoothing) and a correct implementation of the  ISW effect, which will affect the largest scales. However,  the essential intermittent cold spot nature following from  negative $\Phi_{\text{N}}$ will persist. Such cold spots will be polarized just as those deriving from $\zeta_\phi$ are,  and have a CMB spectrum, a prediction for the $\sim 6^{\circ}$ \text{COBE/WMAP} cold spot \cite{coldspot} if it is a smoothed remnant of resonant preheating. 
  
Another non-intermittent {\text{nG}} regime is that of extreme  ``line blending": Fig.~\ref{Fig:numerics}  shows $f_{\chi}\chi_{\text{LF}}^2/M_P^2$ with $f_\chi$= $2\times 10^6$  is a good fit to $\avrg{\fnl\vert \chilf}$ below $\sim \sigma_{\text{HF}}$. (Such a quadratic form also follows from  Eq.~(\ref{fnlchib}).)  For $\chi_{>h} \lsim \sigma_{\text{HF}}$, we will get a power law in $\chi_b$, $\beta_{\chi} \chi_b /M_P+ f_{\chi} \chi_b^2/M_P^2 $, with $\beta_{\chi} $=$2f_{\chi}\chi_{>h}/M_P$. The conventional \text{WMAP} $-9<f_{\text{NL}}<111$ constraints  \cite{coldspot} use $\sim  \zeta_\phi +f_{\text{NL}}\zeta_\phi^2$. In our case, $f_{\chi}\chi^2$ is uncorrelated with  $\zeta_\phi^2$ so the constraint on $f_\chi$ will be considerably relaxed, as long as the linear $\beta_{\chi} \chi_b$ term is subdominant to $\zeta_\phi$, as is expected.  

Further  exploration is needed of how spiked $\fnl$ from resonant preheating may arise  in more general inflation models and on the rich {\text{nG}} impact  and observable constraints of such  $\fnl$ on short and long cosmic scales.  For now, we note SUSY models provide light  non-inflaton fields of the sort we need, and future CMB experiments could test whether {\text{nG}} cold spots are polarized. We thank A.~Chambers, G.~Felder, E.~Komatsu,  A.~Linde and A.~Rajantie for  discussions.

\end{document}